\documentclass[12pt]{iopart}

\usepackage{iopams}  
\usepackage{graphicx}
\usepackage{cite}
\usepackage[colorlinks=true, linkcolor=blue, citecolor=blue]{hyperref}

\begin{document}

\title[Table-top laser-based proton acceleration in nanostructured targets]{Table-top laser-based proton acceleration in nanostructured targets}

\author{M. Blanco$^1$, M.T. Flores-Arias$^1$, C. Ruiz$^2$, M. Vranic$^3$}

\address{$^1$ Departamento de F\'{i}sica Aplicada, Facultade de F\'{i}sica, Universidade de Santiago 
de Compostela, Campus Vida s/n, Santiago de Compostela, Spain}
\address{$^2$ Instituto Universitario de F\'{i}sica Fundamental y Matem\'{a}ticas and Departamento de Did\'{a}ctica de la Matem\'{a}tica y de las ciencias experimentales, 
Universidad de Salamanca, Patio de las Escuelas s/n, Salamanca, Spain}
\address{$^3$ GoLP/Instituto de Plasmas e Fus\~{a}o Nuclear, Instituto Superior T\'{e}cnico, Universidade de Lisboa, 1049-001 Lisbon, Portugal}
\ead{manuel.blanco.fraga@usc.es}
\vspace{10pt}
%\begin{indented}
%\item[]October 2016
%\end{indented}

\begin{abstract}
	
	The interaction of ultrashort, high intensity laser pulses with thin foil targets leads to 
	ion acceleration on the target rear surface. To make this ion source useful for 
	applications, it is important to optimize the transfer of energy from the laser into the 
	accelerated ions. One of the most promising ways to achieve this consists in engineering the 
	target front by introducing periodic nanostructures. 
	
	In this paper, the effect of these structures on ion acceleration is studied analytically 
	and with multi-dimensional particle-in-cell simulations. We assessed the role of the 
	structure shape, size, and the angle of laser incidence for obtaining the efficient energy 
	transfer. Local control of electron trajectories is exploited to maximise the energy 
	delivered into the target. Based on our numerical simulations, we propose a precise range of 
	parameters for fabrication of nanostructured targets, which can increase the energy of the 
	accelerated ions without requiring a higher laser intensity. 
	
\end{abstract}

%\pacs{52.38.Kd, 52.38.-r, 52.65.Rr}
%\maketitle

\section{Introduction}
	
	Ion acceleration in laser-driven plasma accelerators has been a very active field in the 
	last few years. It has been demonstrated that it is possible to accelerate ions up to energies 
	of tens of MeV with table top laser sources \cite{acceleration_Macchi, acceleration_several}. One of the 
	most robust mechanisms used to accelerate ions in plasma based accelerators is the Target 
	Normal Sheath Acceleration (TNSA). In this mechanism a high intensity laser 
	interacts with a few-micron thick solid target to produce energetic ions \cite{TNSA_1, TNSA_2, 
	TNSA_mech, TNSA_proton}. The laser pulse ionizes the target surface and heats up the electrons; 
	these electrons propagate across the target and escape perpendicularly to the rear surface. 
	This generates a space charge separation in the rear surface that yields a strong 
	longitudinal field which can accelerate positively charged particles located in the vicinity 
	of the surface.
	
	Solid targets for plasma-based accelerators can be manufactured with a variety of properties 
	to make the acceleration more efficient. This subject has received a wide attention because the 
	optimization of the targets opens a way to produce more energetic ions without the need of 
	increasing the laser power.
	Different approaches such as varying the target thickness 
	\cite{target_thickness}, nanostructuring the back surface of the target 
	\cite{back_surface, structured_7} or growing a layer of low density foam 
	\cite{foam, foam_2, structured_8} have already been studied. Several 
	publications have reported that adding periodic nanostructures 
	on the target front surface enhances drastically the laser energy absorption 
	\cite{absorption_structured, absorption_structured_2, plasmon_4, structured_2, 
	structured_8, structured_11, andreev, nanoSphere, structured_1}. This generates ions with much higher 
	energies than the ones obtained when targets with a flat surface are used 
	\cite{structured_8, structured_1, structured_3, structured_4, structured_5, structured_6, 
	structured_9, 
	structured_10, structured_12, structured_13, structured_14, nanoSphere}. The 
	nature of this enhancement is still a matter of discussion, however it is known that it is 
	strongly dependent on the shape of the structures, as well as on the angle of incidence of 
	the impinging laser \cite{structured_1, absorption_structured, 
	absorption_structured_2, structured_2, structured_3, structured_10, structured_8, 
	structured_9, structured_11}.
	
	As there are several laser absorption mechanisms in overdense plasmas, such as the 
	generation of Surface Plasma Waves (SPW) \cite{plasmon_1, plasmon_2, plasmon_3, plasmon_4, 
	plasmon_5, plasmon_6, absorption_structured_2, structured_8, structured_13}, resonant absorption, vacuum heating or $J\times B$ heating 
	\cite{Gibbon, heating, electron_heating}, one can use nanostructures with properties especially suited to enhance a 
	particular absorption mechanism. Our aim is to optimize the acceleration of electrons in the 
	vacuum gaps of a periodic structure, so-called vacuum heating 
	\cite{structured_1, structured_2, structured_10}.
	
	The main purpose of this work is to show how the geometry of periodic structures can be 
	optimized to achieve a higher laser energy absorption and proton energies,
	in particular for the specification of the STELA laser of the L2A2 facility of the 
	University of Santiago de Compostela, with laser intensities 
	on the order of $10^{19}~$W/cm$^2$ and a peak power on the order of tens of TW. 
	We present an analytical and numerical study of the 
	interaction of laser pulses and solid targets with triangular periodic nanostructured
	surfaces. The dimensions of the structures dictate the 
	time electrons spend interacting with the laser field
	in the free spaces within the structure which directly influences the ratio of laser energy
	absorbed by the electrons. The reason behind 
	the observed enhanced absorption and higher proton energies is the possibility to
	control the recollision time of the electrons by changing the parameters of the triangular
	structure.
	
	We propose an optimal structured surface for energy absorption and energetic proton production 
	that can be fabricated and used experimentally, providing a robust strategy to obtain higher energy 
	protons without the need of using a higher laser intensity. We introduce a simple analytical model for the
	laser energy absorption by the plasma electrons as a function of the nanostructure dimensions. 
	The predictions of this model are in good agreement with
	our particle-in-cell (PIC) simulation results. 
	In addition, we study the effects of using oblique laser incidence, and propose an alternative 
	structured target surface optimized for oblique incidence.
	
	This paper is structured as follows: Section \ref{sec:II} introduces an analytical model 
	to account for the increase in the energy absorption due to the presence of the periodic 
	nanostructures. Section \ref{sec:III} contains 2D PIC simulations, where a 
	wide range of parameters was studied to find the optimal target design for maximal laser 
	absorption and proton cutoff energy. Section \ref{sec:IV} accounts for differences between 
	2D and 3D geometry. The effect of using oblique incident laser pulses is adressed 
	in Section \ref{sec:V}, in which we propose an optimal experimental configuration to obtain 
	energetic protons. The summary and discussion of our results are presented in Section \ref{sec:VI}.

\section{Analytical model of electron dynamics}\label{sec:II}
	
	In this section we introduce a simple analytical model to understand the electron motion in 
	the presence of the laser field within the periodic structures. The 
	laser energy absorption is directly related to the electron heating, as 
	electrons are the lightest particles in the plasma and the first to interact with the laser
	field. Understanding how electrons absorb the laser energy and carry it towards 
	the rear of the target is crucial to select an optimal surface structure. 
	
\begin{figure}[ht!]
		\centering
			\includegraphics[width=128mm]{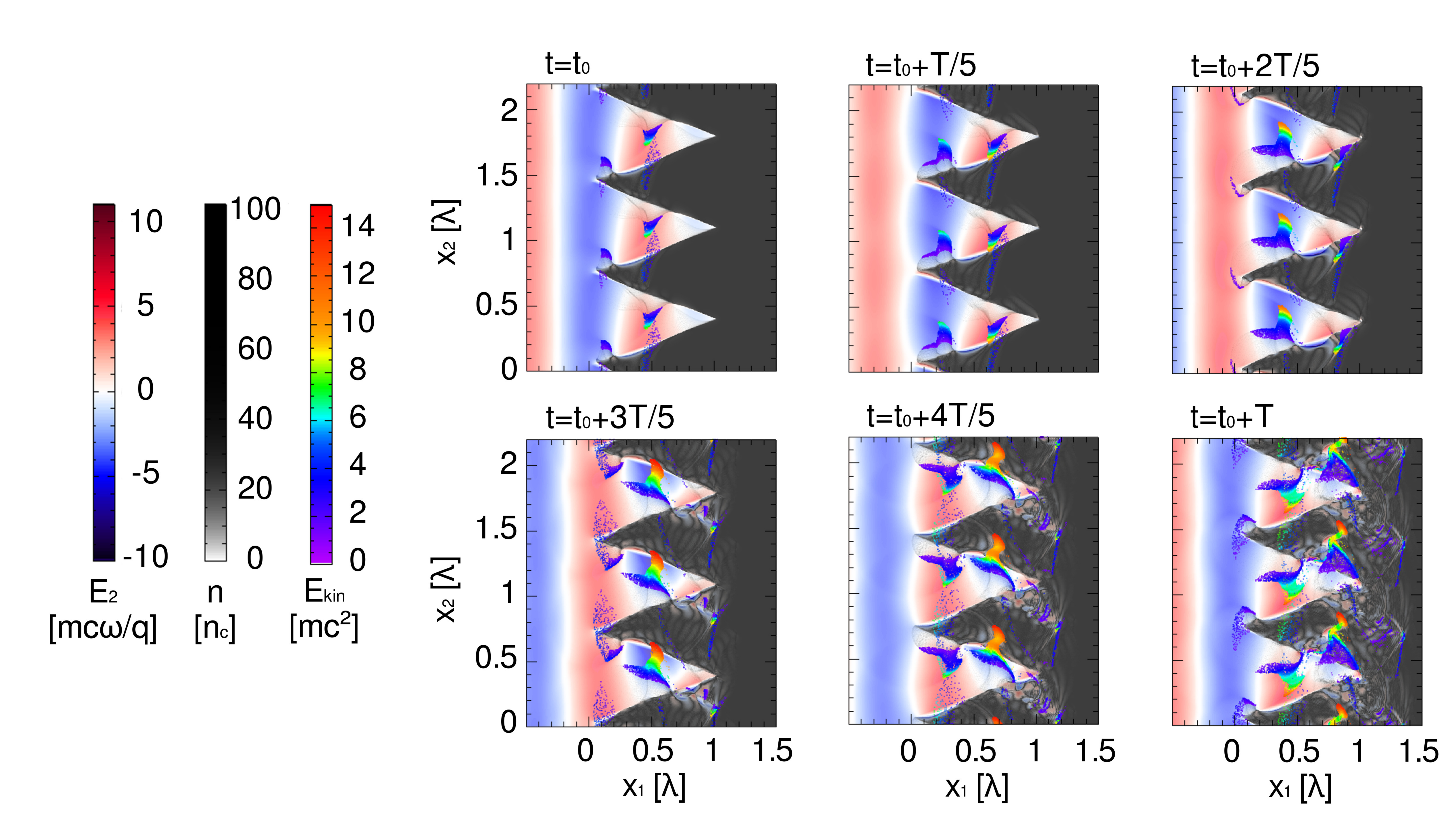}
		\caption{Illustration of the electron dynamics over a laser period $T$ 
		with six snapshots of the electron density, the $E_2$ component of the electric 
		field, in which the laser field is polarized, and the most energetic electrons, 
		coloured depending on their kinetic energy. The electrons move in the vacuum gap 
		from one triangle to the next according the sign of the $E_2$ component of the 
		electromagnetic field, gaining energy in the process.}
		\label{fig:figure1}
	\end{figure}

	The use of triangular structures in the front face of the target changes the local angle of 
	incidence of the laser and allows electrons to undergo a temporary interaction with the 
	laser field in vacuum before recolliding with the target. As the laser arrives to the 
	target, it strips off a portion of the electrons from the lateral surface of the structure. 
	The electrons are accelerated by the laser field in the vacuum gaps of the structures and 
	gain energy. They reenter the target where the laser cannot penetrate and deliver their 
	energy into the target, as shown in figure \ref{fig:figure1}. The shape of the triangular 
	structures determines the time the electrons spend being accelerated  in the vacuum gap and 
	their recollision energy. We expect an optimal laser absorption for targets with geometrical 
	properties that allow for maximum electron energy gain.
	
	A simple model that neglects all fields except the laser is useful to understand how the 
	recollision energy is related to the triangle shape.
	Our starting point is to consider the relativistic motion of an electron in vacuum 
	under the influence of a linearly polarized electromagnetic wave given by the vector 
	potential $\vec{A} = a_0 \frac{m_e c}{\omega} \sin(\varphi) ~\hat{x}_2$, where $a_0$ is the 
	dimensionless vector potential defined by $a_0 = 0.85 \left(
	\frac{I \lambda^2}{10^{18} W cm^{-2} \mu m^2}\right)^{\frac{1}{2}}$, where $I$ is the laser intensity in
	Wcm${}^{-2}$ and $\lambda$ the laser wavelength in $\mu$m. 
	The variable $\varphi = \omega t - \vec{k} \cdot \vec{r} + \varphi_0$ represents the electromagnetic wave phase, $\vec{k}$ is the wave vector, $\vec{r}$ the particle position, 
	$\omega$ the angular frequency, $t$ the time and $\varphi_0$ is the initial phase. The 
	equation of motion of an electron in the vacuum gap is determined by the Lorentz force:
	
	\begin{equation}
	 \frac{d\vec{p}}{dt} = -q_e\vec{E} - q_e\left(\frac{\vec{p}}{\gamma m c} \times \vec{B}\right) = 
	 \frac{q_e}{c}\frac{\partial \vec{A}}{\partial t} - q_e\left(\frac{\vec{p}}{\gamma m c} \times 
	 \left(\vec{\nabla} \times \vec{A}\right)\right)
	\end{equation}
	
	\noindent where $q_e$ is the electron charge. This equation can be solved for the specified 
	vector potential assuming that the electron is initially at rest and that for $t = 0$ we have $\varphi = 0$. The momentum 
	and displacement of the electron are then given by \cite{Gibbon}:
	
	\begin{eqnarray} \label{eq:002a}
		p_1 = \frac{a_0^2}{2} m_e c \sin^2(\varphi) & \quad p_2 = a_0 m_e c \sin(\varphi) 
		\\ \label{eq:002b}
		\Delta x_1 = \frac{a_0^2}{8 \pi} \lambda 
		\left(\varphi - \frac{\sin(2 \varphi)}{2} \right) & \quad \Delta x_2 = 
		\frac{a_0}{\pi} \lambda \sin^2 \left(\frac{\varphi}{2}\right)
	\end{eqnarray}

	\noindent where the indexes ``1'' and ``2'' refer to the longitudinal and transverse field
	directions and $\lambda = 2 \pi c / \omega$ is the laser field wavelength.
	
	The maximum electron energy is reached when $\varphi = \pi / 2$, because both components of 
	the electron momentum are maximized, so if the electrons reenter the target at this point 
	they will absorb the maximum possible energy from the field. Reaching the optimal 
	phase at the moment when the electron reenters the target is controlled by its initial
	position in the structured surface, expressed by its initial height, $h_0$. We can establish 
	a relation between the initial height where a single electron is located and the 
	phase when it arrives to the surface of the next triangle by 
	using the displacements in equation (\ref{eq:002b}). 
	If $h$ and $w$ are the structure height and width respectively, this relation is given by:
	
	\begin{equation} \label{eq:003}
		h_0(\varphi) = \frac{a_0^2}{16 \pi} \lambda \left(\varphi - \frac{\sin(2 \varphi)}{2}\right) 
		+ \frac{a_0}{\pi} \frac{h}{w} \lambda \sin^2 \left(\frac{\varphi}{2}\right) 
	\end{equation}
	
	The maximum amount of energy absorbed by the electrons in the gap is reached when 
	$h_0(\varphi = \pi / 2) \approx h$, because this means that the electrons initially 
	located at the tip of the structure gain the maximum 
	possible energy. These electrons are the first to interact with the laser 
	pulse. If $h_0(\varphi = \pi / 2) > h$ the electrons will not reach the 
	maximum energy because their ideal initial height is not allowed on the structure and when 
	$h_0(\varphi = \pi / 2) < h$ a portion of the electrons at the top of the 
	structure will stay longer in the vacuum gaps and not enter the target with the optimal 
	energy. It can be also noted that 
	the second term on the right hand side of equation (\ref{eq:003}) becomes less relevant as 
	$a_0$ increases, which suggests that the effect of the structures becomes less important as 
	the field amplitude increases. Using a laser field with a dimensionless amplitude of 
	$a_0 = 4$, which can be achieved with the laser STELA, the equation (\ref{eq:003}) for the maximum 
	electron energy becomes:
	
	\begin{equation}\label{eq:004}
		h_0(\varphi = \pi / 2) = \frac{\lambda}{2} + \frac{2}{\pi} \frac{\lambda}{w} h
	\end{equation}
	
	The value of $h_0$ relative to $h$ in the previous equation is controlled by the width of 
	the structures. For $w \ll 2 \lambda / \pi$ then $h_0 \gg h$ and viceversa. Therefore, there 
	is an optimal structure width for improving the energy absorption, given by 
	$w = 0.64\lambda$. When the structure width has this value, equation (\ref{eq:004}) becomes 
	$h_0(\varphi = \pi / 2) = \lambda/2 + h$, which means that for higher structures $(h \gg \lambda/2)$ we get
	$h_0 \rightarrow h$. We therefore expect that above a certain structure height 
	$(h \sim 0.5 \lambda)$ the energy absorption percentage reaches a maximum value and does not 
	change further. This simple model provides a clear picture on how the engineering 
	of the triangular structures can be used to control the electron trajectories and maximize 
	the energy they deliver to the target.
	
\begin{figure}[ht!]
		\centering
			\includegraphics[width=138mm]{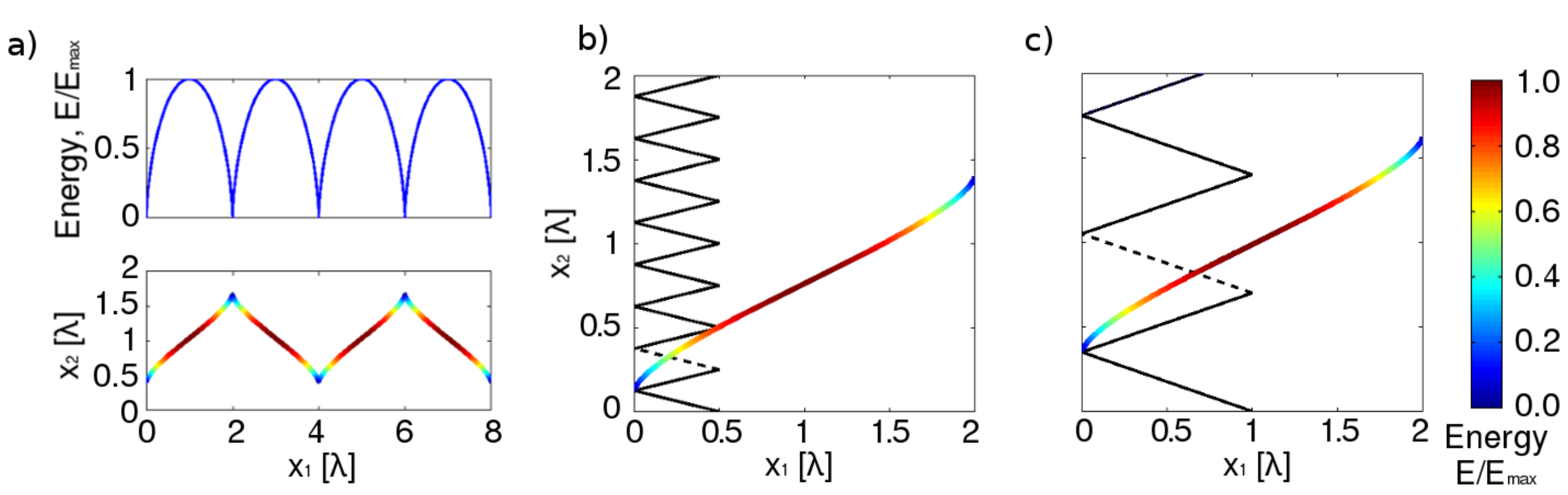}
		\caption{(a) Energy of one electron versus the longitudinal coordinate $x_1$ and its
		trajectory. The trajectory is colored according to its energy. Electron trajectories within the vacuum gaps colored according to their energy for 
		different structure heights and widths: (b) $h = 0.5 \lambda$ and $w = 0.25\lambda$, 
		and (c) $h = \lambda$ and $w = 0.7\lambda$. The dotted line represents the 
		surface where the electron reenters the target.
		All panels are for a dimensionless vector potential of $a_0 = 4$.}
		\label{fig:figure2a_2c}
	\end{figure}

	Figure \ref{fig:figure2a_2c} depicts the energy and motion of an electron under equations
	(\ref{eq:002a}) and (\ref{eq:002b}) and shows that the structure shape can be manipulated to 
	obtain maximum absorption from the electrons moving from the tip of one triangle across the 
	vacuum gap. Figure \ref{fig:figure2a_2c}a shows that there
	is a maximum energy, reached periodically for certain positions that would correspond to the phase
	$\varphi = (2N + 1) \pi / 2$, where $N$ is an integer. Figures \ref{fig:figure2a_2c}b-c show, 
	for different structure 
	heights and widths, the trajectory of an electron moving in the vacuum gap, 
	colored according to its energy. It can be observed that the shape of the structure 
	influences the energy that the electron has at the recollision time; the energy is lower in 
	figure \ref{fig:figure2a_2c}b than in \ref{fig:figure2a_2c}c.
	
	The predictions obtained from the model for the dependence of the energy absorption with the 
	structure height and width can be tested by performing PIC simulations. In the following 
	sections we discuss the results obtained from such simulations, with the aim of designing 
	an optimal target for energy absorption and ion acceleration.

\section{Effect of the structure shape and size on the laser energy absorption and proton 
acceleration}\label{sec:III}

\begin{figure}[ht!]
		\centering
			\includegraphics[width=133mm]{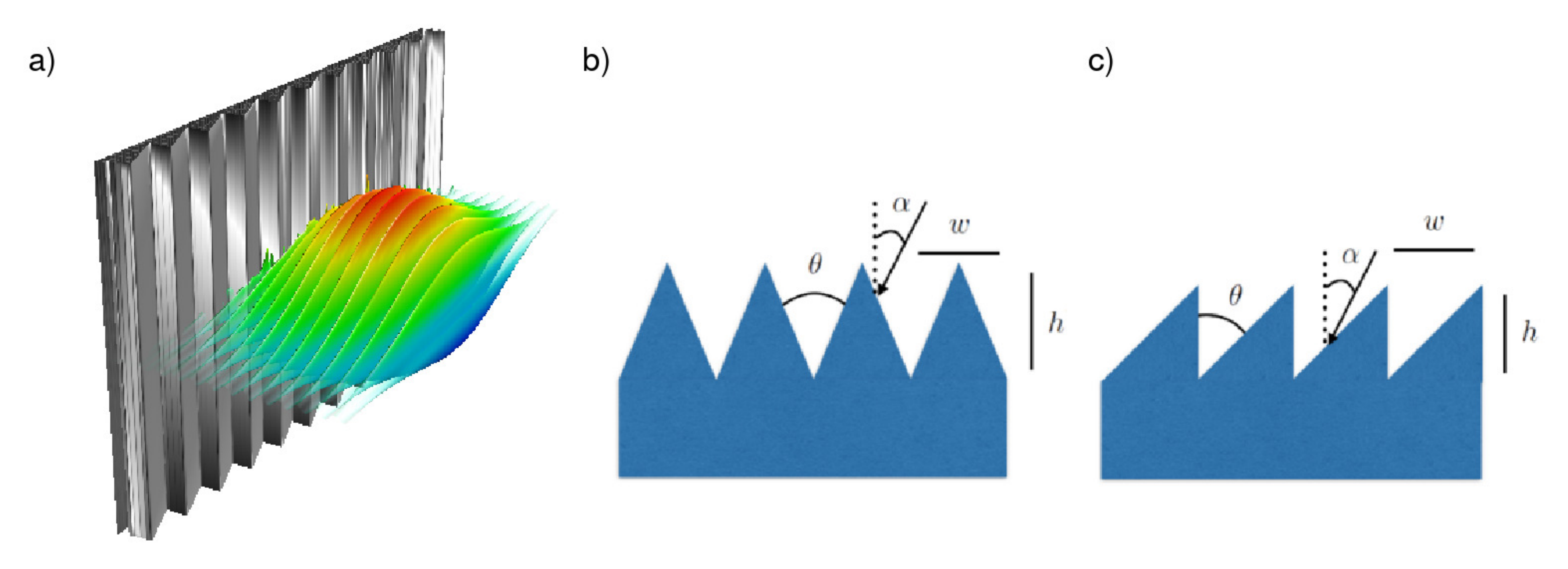}
		\caption{Shapes of the triangular structures. (a) 3D image of the simulation setup 
		and (b) - (c) detailed view of the 2D targets. The variable $h$ represents structure height, $w$ the width, $\theta$ the angle formed between two neighbouring structures and $\alpha$ the angle of lase incidence towards the target.}
		\label{fig:figure3a_3c}
	\end{figure}

	The aim of this section is to identify a parameter range with maximum transfer of laser 
	energy to the accelerated protons. To adress how the shape of the structures affects the 
	absorption of laser energy as well as the energies of the electrons and protons, we present 
	a numerical study encompassing a wide range of sizes for two types of 
	structures indicated in figure 
	\ref{fig:figure3a_3c}. We vary their width and height and use a laser pulse at normal incidence.
	
	Numerical simulations of the laser-plasma interaction are performed with the particle-in-cell 
	(PIC) code OSIRIS \cite{osiris}. In OSIRIS, the fields are stored on a discretized spatial 
	grid and advanced according to the Maxwell's equations. The particle motion is determined by 
	the relativistic Lorentz force. 
	
	Two different targets with triangular structures were used in this work, as shown in 
	figure \ref{fig:figure3a_3c}. The reasons behind choosing these structures are that they yield 
	high absorption rates and efficient proton acceleration in comparison with other 
	kinds of structures \cite{structured_3, structured_11} and can be manufactured for their use 
	in experiments. The assimetry presented in 
	figure \ref{fig:figure3a_3c}c (``tilted triangles'') with respect to figure \ref{fig:figure3a_3c}b
	(``regular triangles'') is interesting from the experimental point of view, where oblique 
	incident laser pulses are going to be used.
	
	The targets are made of electrons and protons with a number density of $n = 40 n_c$, where 
	$n_c = \frac{m_e \omega^2}{4\pi q_e^2}$ is the critical plasma density. All the targets 
	have a bulk thickness of $0.5 \lambda$ (unless specified differently), where $\lambda$ is
	the wavelength of the laser, and the number of particles per cell is $16$ per species. 
	The density has a steep profile as we consider a high contrast laser 
	$(> 10^{10} ~$at$~ 5 ~$ps$)$ which corresponds to the STELA laser.
	The simulation box has a width and length of 
	$38.3 ~\mu$m and $16.9 ~\mu$m, respectively. The spatial resolution is 
	$\delta = 2.55 ~$nm in both axes. 
	
	The laser pulse is focused on the target surface. This pulse is launched from the left 
	wall of the simulation box, located at a distance of $8.9 ~\mu$m to the target. The laser 
	has an intensity of $3.45 \times 10^{19} ~$W/cm$^2$, a FWHM of $25 ~$fs 
	(with a $sin^2$ temporal profile), a wavelength of $\lambda = 800 ~$nm and a 
	spot diameter of $6 ~\mu$m. The laser pulse is linearly polarized in the simulation 
	plane, such that it is always p-polarized in relation to the structures.
	
	The simulation advances in timesteps of $4.26 ~$as. The reflected energy, as well as the 
	electron properties at the rear surface, are measured right after the interaction finishes, 
	at the time $t = 70.2 ~$fs. The proton properties are measured at $t = 172.3 ~$fs, the 
	time at which electrons that generate the accelerating field start leaving the simulation 
	box. The electron temperature at the rear surface is obtained by fitting the electron 
	spectrum to a Maxwell-J\"{u}ttner distribution. The simulation setup is 
	designed to scan the parameter space and compare the relative gain between the flat and structured targets.
		
\begin{figure}[ht!]
		\centering
			\includegraphics[width=146mm]{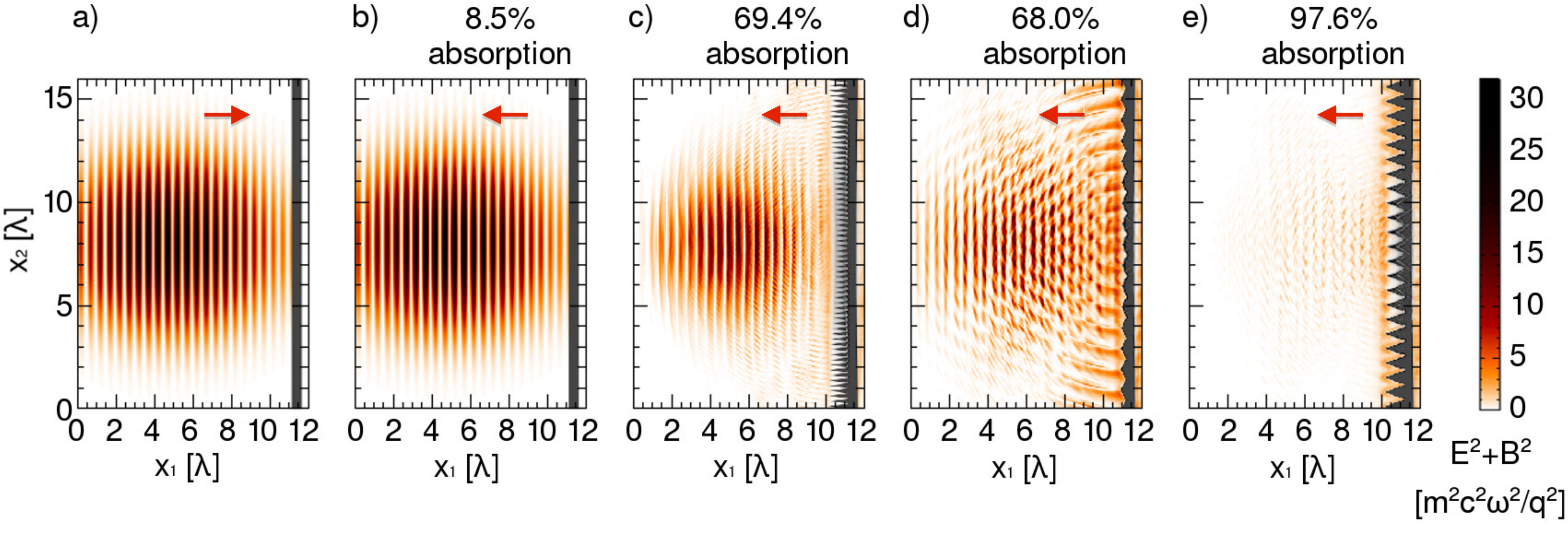}
		\caption{(a) Incident and (b)-(e) reflected electromagnetic energy density for 
		different structured targets. The reflected field is shown for a (b) flat target and 
		targets with structure height and width of: (c) $h = \lambda$ and $w = 0.25\lambda$, 
		(d) $h = 0.25\lambda$ and $w = \lambda$ and (e) $h = \lambda$ and $w = 0.7\lambda$.}
		\label{fig:figure4a_4e}
	\end{figure}

	Several cases from low to almost complete laser absorption are illustrated in Figure \ref{fig:figure4a_4e}.
	Figure \ref{fig:figure4a_4e}a shows the incident laser pulse and figures
	\ref{fig:figure4a_4e}b-e display the reflected fields. The lowest absorption is obtained for a flat target. The results obtained 
	for different structured targets are displayed in figures \ref{fig:figure4a_4e}c-e. We observe that the reflected spatial distribution of 
	light when structured targets are used carries the imprint of the nanostructures at the target surface, and more laser energy is absorbed.
	
\begin{figure}[ht!]
		\centering
			\includegraphics[width=140mm]{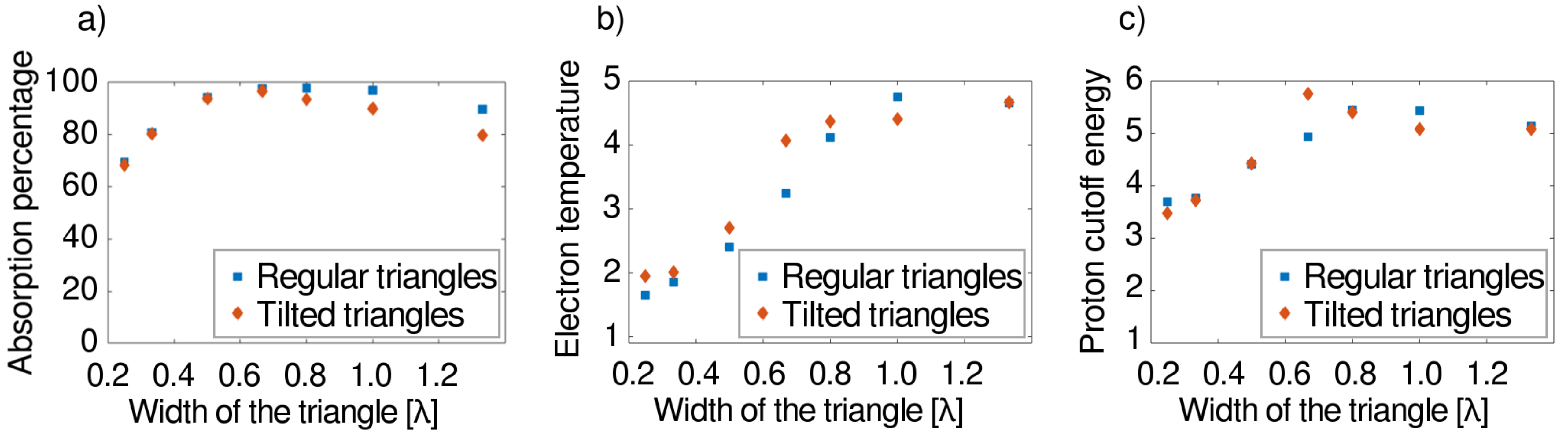}
		\caption{(a) Laser energy absorption 
		percentage, (b) electron temperature and (c) proton cutoff energy as a function of the structure width. The height 
		of the structures is $h = \lambda$. The electron temperature and proton energy 
		are normalized to the values obtained for a flat target.}
		\label{fig:figure5a_5c}
	\end{figure}

	To verify the predictions of Section \ref{sec:II}, we perform a first set of simulations 
	with a fixed structure height of $h=\lambda$, where we vary the structure width. According 
	to our analysis 
	we should expect maximum energy absorption at a width of $w \sim 0.64\lambda$. Figure \ref{fig:figure5a_5c}a shows the energy absorption, as a function of the structure width, for regular and tilted triangular nanostructures.  In both cases 
	the energy absorption increases with the width of the triangles up to a 
	maximum nanostructure width around $0.7 \lambda$. Above this value the energy absorption decreases 
	smoothly as the triangle width increases. For the triangles $0.6\lambda~-~\lambda$ wide, 
	we obtain that the energy absorption is above $90\%$. The increase in
	electron temperature is shown in figure \ref{fig:figure5a_5c}b: it
	rises with the width of the triangles for both types of structures. 
	Once the width of the triangles achieve a value close 
	to $0.7\lambda$ the slope of the curve changes to a lower value.
	The gain in proton cutoff energy is shown in \ref{fig:figure5a_5c}c. It exibits a similar 
	trend as the energy absorption.
	
	The plots in figure \ref{fig:figure5a_5c} show that the the structure width 
	can be optimized to yield a maximum laser absorption and proton energy cutoff. The 
	structure width of the optimal target is consistent with the analytically predicted $w = 0.64 \lambda$. 
	
\begin{figure}[ht!]
		\centering
			\includegraphics[width=140mm]{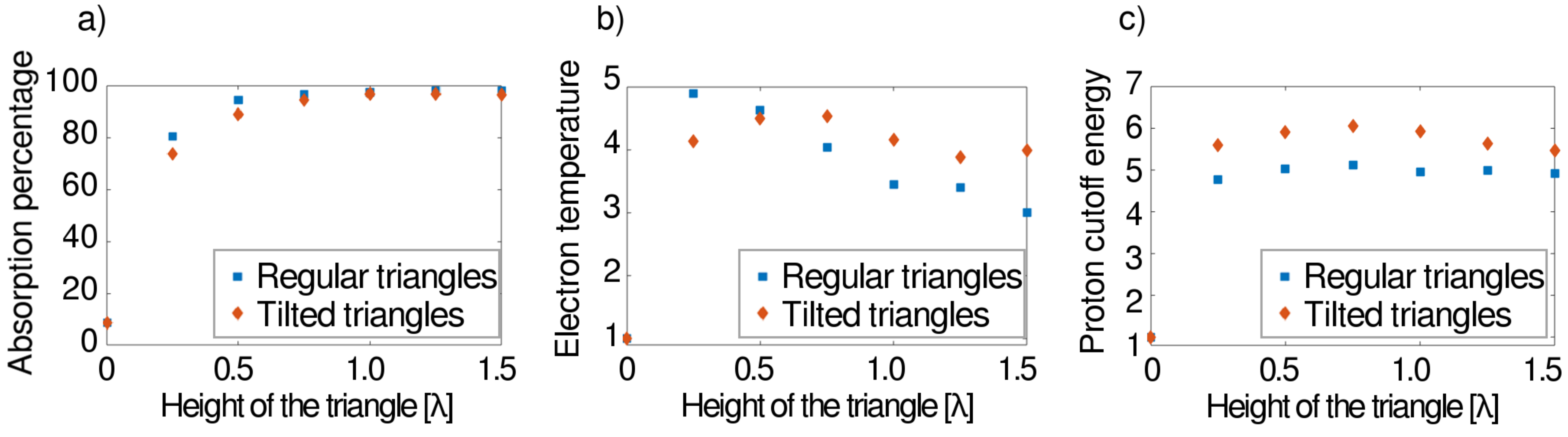}
		\caption{(a) Laser energy absorption 
		percentage, (b) electron temperature and (c) proton cutoff energy as a function of a structure height. 
		The width of the structures is $w = 0.7 \lambda$. The electron temperature and 
		proton energy are normalized to the values obtained for a flat target.}
		\label{fig:figure6a_6c}
	\end{figure}

	A second set of simulations is performed with a fixed structure width of $w = 0.7 
	\lambda$ and varying the structure height. We kept the rest of simulation parameters equal 
	to the ones in the previous case. Here we should expect nearly a complete laser energy 
	absorption above a certain threshold structure height. Figure
	\ref{fig:figure6a_6c} displays the energy absorption, the relative electron 
	temperature and proton 
	cutoff energy versus the height of the structures. The energy absorption percentage 
	is shown in figure \ref{fig:figure6a_6c}a, that depicts an increase of the absorption of 
	energy as the height of the triangles becomes bigger, up to a 
	maximum value, close to $100\%$. Once this maximum is achieved, it remains unchanged as we increase the 
	height of the triangles. The electron temperature, in figure 
	\ref{fig:figure6a_6c}b, shows a big increase when the structures are added, 
	followed by a smooth decay/stabilization for higher structures. Figure 
	\ref{fig:figure6a_6c}c displays the relative proton cutoff energy.
	
	For $h > 0.5 \lambda$, the results in figure \ref{fig:figure6a_6c} show that the 
	absorption percentage is above $90 \%$ and there is a 5-fold increase in the proton cutoff 
	energy compared with the flat target. The high absorption percentages 
	shown in figure \ref{fig:figure6a_6c}a are due 
	to the choice of a structure width of $0.7\lambda$, close to the optimal value found before.
	As predicted by the analytical model, above the threshold structure height, the laser energy 
	absorption is nearly $100\%$ and there is no significant difference observed in the spectrum 
	of generated protons.
	
	Figures \ref{fig:figure5a_5c} and \ref{fig:figure6a_6c} demonstrate that there is a 
	correspondence between the trend followed by the laser absorption and the cutoff energy of 
	the protons. This is not surprising, because in TNSA the absorbed laser energy is carried by 
	the electrons towards the rear side of the target. These electrons escape the target and 
	create a longitudinal field proportional to the square root of the electron temperature and 
	to the electron front number density \cite{TNSA_mech}. This longitudinal field is 
	responsible for the acceleration of protons. We therefore expect higher proton energies for
	higher achieved electron temperatures at the rear surface. However, additional height in the 
	structures changes the effective target thickness and hence the electron front number 
	density is also modified. The consequence is that the electrons with a lower temperature 
	(e. g. for $h=1.5 \lambda$ in figure \ref{fig:figure6a_6c}b) can, in principle, generate the 
	accelerating field of the same magnitude as the electrons with a $53\%$ higher temperature 
	in a case with a different effective target thickness (e. g. for $h=0.5\lambda$ in figure 
	\ref{fig:figure6a_6c}b). We therefore obtain similar values for the accelerating field and 
	for the proton cutoff energy in all cases where $h>0.5 \lambda$. 
	
\section{Comparison between 2D and 3D results} \label{sec:IV}
	
	We have analyzed how the energy absorption varies when triangular structures are 
	on the target surface and how this variation affects the electron heating and 
	consequently the energies of the accelerated protons. Our theoretical analysis, combined 
	with 2D PIC simulations, shows that the structure shape can be optimized to yield high 
	percentages of energy absorption, which is in agreement with previous publications
	\cite{structured_1, structured_2, structured_10}.
	
	The values obtained for the optimal height and width in terms of energy absorption are given 
	by a width of $w = 512 ~$nm and a lower bound for the structure height of 
	$h \sim 400 ~$nm. Fabrication of these structures is achievable by current techniques, 
	therefore the targets described here can be manufactured in a cost effective way and used 
	experimentally \cite{structured_3, structured_4, structured_8, structured_11}.
	
	Our conclusions regarding the target structure for optimal laser absorption are general and 
	can be extended to 3D geometry. The electron interaction with an electromagnetic wave in 
	vacuum is fully described in 2D. The energy of the electrons at the point of re-entry into 
	the target depends on the geometrical properties of the structures such as width and height 
	and it is intrinsically a 2D problem. The optimal configuration for laser absorption is 
	therefore likely to be the same in 2D and 3D geometry. However, even though ion acceleration 
	can be studied qualitatively in 2D, it is well-known that the proton energy cutoff in 
	TNSA is lower in 3D geometry \cite{2D3D, 2D3D_2}. 
	
	Due to limited computational resources, it is not possible to perform a full-scale 3D 
	simulation and allow enough time for target expansion. However, the 
	3D simulations can be performed in slab geometry. In this geometry the laser is treated as a 
	wavepacket that is transversely a plane wave and periodic transverse boundary conditions are applied 
	both for the fields and the particles, except for the direction of laser propagation where 
	open boundaries are used.
	
	The simulation box is $25 \lambda$ long and $3.5 \lambda \times 3.5 \lambda$  wide. The 
	laser pulse is initialized inside the box with a total duration of $7 T$, where $T = c / \lambda$ is the laser period, and a 
	distance of $0.5\lambda$ to the structure tip. We performed simulations for optimal 
	structured targets where $h = \lambda$ and $w = 0.7\lambda$ and for flat targets, with a goal to compare the relative increase of the absorption efficiency in 2D and 3D.
	
	In 3D simulations we obtained $91.5 \%$ of laser absorption for regular, $89.5 \%$ for tilted triangles and $2\%$ for a flat target. In 2D simulations, the $95.4\%$ was absorbed for regular, $92.3\%$ for tilted triangles, and $2.6\%$ for the flat target. The absorption estimates from 2D and 3D are consistent within a margin of $4\%$. There is approximately a 15-fold increase in the proton cutoff energy when we use structured targets instead of flat ones in 3D. 
	
\begin{figure}[ht!]
		\centering
			\includegraphics[width=120mm]{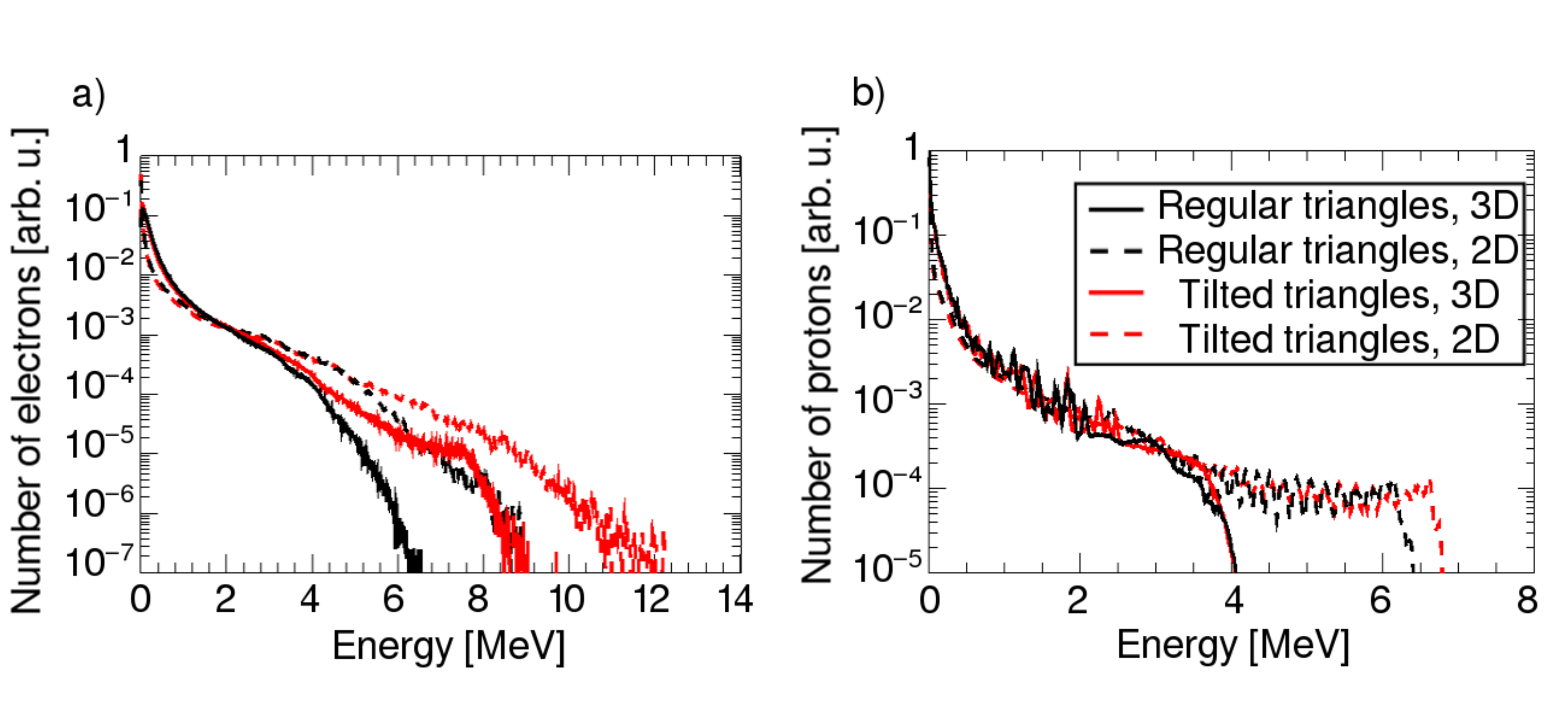}
		\caption{The spectrum of (a) electrons and (b) protons measured 
		for the 2D and 3D setup. The height and width of the structures are $h = \lambda$ and 
		$w = 0.7 \lambda$.}
		\label{fig:figure7a_7b}
	\end{figure}

	The electron spectra from 2D and 3D simulations at the time $t=25.5~$fs are displayed in 
	figure \ref{fig:figure7a_7b}a. They are normalised to the same reference height at the 
	energy $2$MeV. We can observe a slight difference between the tilted and regular structures,
	due to different electron dynamics at the target surface. 
	The respective proton spectra at the time $t = 51.1~$fs are shown in figure 
	\ref{fig:figure7a_7b}b. These spectra are normalised to the same reference height at 
	the energy $1$MeV. As expected, the proton cutoff is lower in 3D compared to the 2D case. 
	Apart from verifying the conclusions obtained in 2D, an additional advantage of 3D 
	simulations is that they can provide an estimate of the total number of accelerated protons. 
	In our case the number of protons being accelerated to an energy above $0.1$MeV is 
	approximately $1.35 \times 10^{11}$, corresponding to over $\sim 21.6~$nC of 
	charge accelerated to energies up to $4~$MeV. As the transverse box dimensions are 
	chosen to be on the order of the laser spotsize, this is the approximate number of protons 
	per shot expected in an experiment with similar conditions.
	
\section{Oblique laser incidence}\label{sec:V}

	In the previous sections it has been discussed how the energy absorption and the particle 
	properties change with the front
	structure shape of the solid target. All the previous simulations have been performed with a 
	pulse in normal incidence, however experiments of TNSA proton acceleration are typically done 
	in oblique incidence. The main reason for using oblique incidence is to avoid 
	the damage on the optical elements used to transport the beam to the target, with the 
	particles ejected by the target or the back reflection of the laser pulse.
	
	The angle of incidence of the laser pulse is expected to affect the laser energy 
	absorption and the particle energies. We have performed simulations at different angles of incidence using targets with $w = \lambda$ and $h = 1.2 \lambda$, for regular and $w = \lambda$ and $h = \lambda$ for tilted triangles.
	
\begin{figure}[ht!]
		\centering
			\includegraphics[width=140mm]{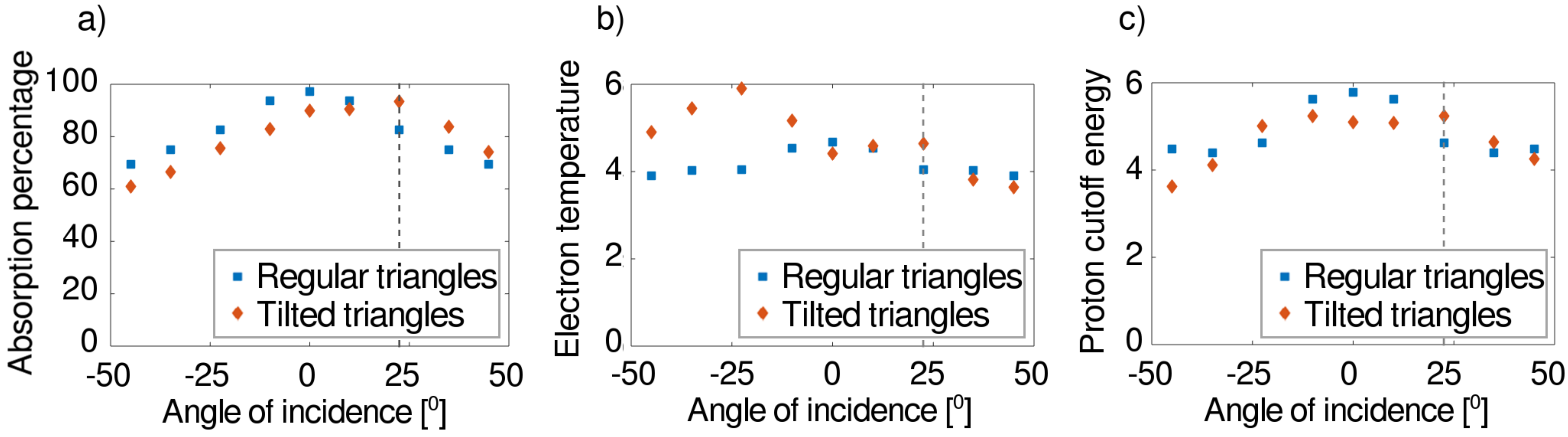}
		\caption{(a) Laser energy absorption percentage, (b) electron temperature and 
		(c) proton cutoff energy. The grey dotted line indicates the angle of incidence 
		$\alpha = 22.5^o$, where the maximum energy absorption for the tilted structure 
		occurs. The width and height of the 
		structures are $w = \lambda$ and $h = \lambda$ for the tilted 
		triangular structure and $h = 1.2\lambda$ for the regular triangular structure. The values are 
		normalised to the results obtained for a flat target and a normally incident laser 
		pulse.}
		\label{fig:figure8a_8c}
	\end{figure}

	Figure \ref{fig:figure8a_8c} displays 
	(a) the energy absorption, (b) the electron temperature and (c) the proton cutoff energy at oblique laser incidence. 
	Figure \ref{fig:figure8a_8c}a shows 
	that for the regular triangular structure, the maximum absorption and particle energies occur at normal 
	incidence, while for the tilted structures this maximum is shifted by a value close to 
	$22.5^o$. The electron temperature for the tilted structures, in 
	\ref{fig:figure8a_8c}b, displays two maxima at the angles $\pm 22.5^o$, while the
	proton cutoff energy for the same structure, in \ref{fig:figure8a_8c}c, shows a plateau around 
	the zero angle slightly maximized for the maximum energy absorption angle.
	
	The peak observed for the elecron temperature
	at $\alpha = -22.5^o$ represents electrons that get heated efficiently but do not contribute 
	to enhance the energy of TNSA protons. This behaviour suggests that a surface plasma wave 
	(SPW) \cite{plasmon_1, plasmon_2, plasmon_3, plasmon_4, plasmon_5, plasmon_6, structured_13} 
	is being excited at this angle, such that the electrons escape the target tangentially. This 
	is confirmed by measuring the longitudinal accelerating field in the rear surface of the 
	target, which shows the same trend as the proton cutoff energy in figure 
	\ref{fig:figure8a_8c}c.
	
	The analysis of the effect of oblique incident pulses shows that the assymmetric tilted 
	triangles yield to a higher absorption percentage and more energetic protons at oblique 
	incidence compared to normal incidence. We have performed an additional simulation combining 
	an oblique incident laser pulse with an optimized height and width 
	of the structure $h=\lambda$ and $w=0.7\lambda$. The bulk target thickness is $2\lambda$ 
	and the angle of incidence of the laser is $\alpha = 17.5^o$. 
	
\begin{figure}[ht!]
		\centering
			\includegraphics[width=135mm]{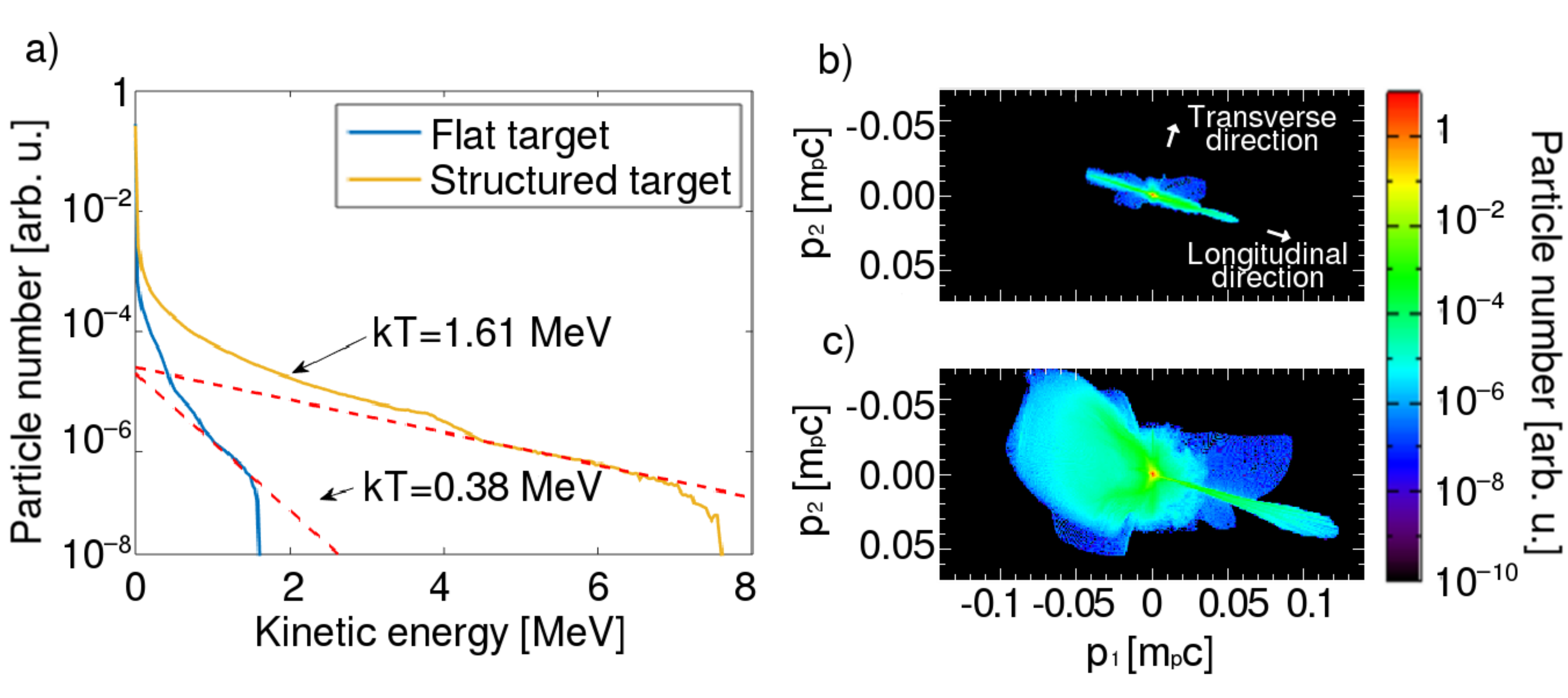}
		\caption{(a) Proton spectrum and the p1-p2 momentum space for (b) a flat and (c) a structured target. }
		\label{fig:figure9a_9c}
	\end{figure}

	The comparison between the results obtained with the same setup for a target with a flat 
	surface and the structured target show a 15-fold increase in the energy absorption 
	percentage, from a $6.1\%$ in the flat target to a $90.6\%$ in the structured target. 
	This enhancement of the energy absorption generates protons with energies between 4 and 5 
	times higher compared with the case in which a flat solid target was used.
	This is verified in Figure \ref{fig:figure9a_9c} that shows (a) the proton energy spectrum and the momentum space $p_1-p_2$ of the protons. Both, the cutoff energy and the temperature of the 
	protons increase by a factor between 4 and 5 when one uses a structured target in place of a flat one. 
	
	The result shown in figure \ref{fig:figure9a_9c} demonstrates in an experimentally feasible setup that 
	the use of surface nanostructured targets can increase substantially the 
	energy of the accelerated protons. Our simulated targets are made of electrons and protons, 
	but real targets are composed of heavy ions and electrons with a thin proton contamination 
	layer on the rear surface \cite{acceleration_Macchi, TNSA_1, TNSA_proton}, as in the typical 
	TNSA scenario, hence in an experiment we should expect different proton energy cutoffs. 
	However, the dynamics of 
	the surface electrons will be similar as in our simulations, because their motion is 
	controlled mainly by the laser field and the shape of the structures, this means that the 
	absorption percentage will be similar to what we have obtained in our results.
	We have performed a set simulations with targets made of 
	heavy ions and electrons and our results were confirmed. The optimal structure for energy 
	absorption is still the same, however the energies of accelerated protons and ions change, because the charge separation field at the target rear will be different and 
	heavier ions will move with different velocities.
	
\section{Conclusions}\label{sec:VI}
	
	The acceleration of protons in laser plasma accelerators using thin targets with periodic 
	nanostructures has been studied. We have performed 2D and 3D PIC simulations
	to reveal how the energy absorption and the accelerated proton kinetic energies vary with
	the shape of the structures and the angle of incidence of the laser. 
	
	The use of triangular periodic structures at the target front surface increases 
	the laser energy absorption and the proton kinetic energy. Through an analytical and 
	numerical study, we have found that these quantities can be maximized by tailoring the 
	dimensions of the nanostructures. 
	
	Through the right choice of the dimensions of the structures and the angle of incidence of 
	the laser, energy absorption percentages on the order of $90\%$ can be achieved, 
	yielding to an increase on the proton kinetic energy between 4 and 5 times in comparison 
	to those that can be achieved with flat targets. The structured target optimizes the 
	laser absorption by the electrons. The absorption mechanism is independent of the ion charge in the target,
	density or thickness, and therefore it is expected to apply for a wide variety of materials.
	
	Our results show that by using nanostructured targets one can obtain energetic ions with a 
	commercially available table-top laser system. 
	This constitutes a robust strategy to produce high energy protons for applications with 
	these table top lasers.

	\ack
	
	This work has been partially supported by the Xunta de Galicia/FEDER under contract Agrup2015/11 (PC034) and by MINECO under contract FIS2015-71933-REDT.
	The authors would like to acknowledge the OSIRIS Consortium, consisting of UCLA 
	and IST (Lisbon, Portugal) for the use of OSIRIS, for providing access to the OSIRIS 
	framework. M. Blanco also thanks the Ministry of Education of the Spanish government for 
	the FPU fellowship. Camilo Ruiz also thanks MINECO project 
	FIS2016-75652-P. M. Vranic acknowledges the support of ERC-2010-AdG Grant 267841 and LASERLAB-EUROPE IV - GA No. 654148.
	Simulations were performed at the Accelerates cluster (Lisbon, Portugal).
	
	\section*{References}
	
\bibliographystyle{unsrt}
\bibliography{biblio}

\begin{thebibliography}{10}

\bibitem{acceleration_Macchi}
A.~Macchi, M.~Borghesi, and M.~Passoni.
\newblock {\em Rev. Mod. Phys.}, 85:751--793, 2013.

\bibitem{acceleration_several}
A.~Stockem~Novo, M.~C. Kaluza, R.~A. Fonseca, and L.~O. Silva.
\newblock {\em Sci. Rep.}, 6:29402, 2016.

\bibitem{TNSA_1}
S.~C. Wilks, A.~B. Langdon, T.~E. Cowan, M.~Roth, M.~Singh, S.~Hatchett, M.~H.
  Key, D.~Pennington, A.~MacKinnon, and R.~A. Snavely.
\newblock {\em Phys. Plasmas}, 8(2):542--549, 2001.

\bibitem{TNSA_2}
R.~A. Snavely et~al.
\newblock {\em Phys. Rev. Lett.}, 85:2945--2948, 2000.

\bibitem{TNSA_mech}
P.~Mora.
\newblock {\em Phys. Rev. Lett.}, 90:185002, 2003.

\bibitem{TNSA_proton}
J.~Fuchs et~al.
\newblock {\em Phys. Rev. Lett.}, 94:045004, 2005.

\bibitem{target_thickness}
A.~J. Mackinnon, Y.~Sentoku, P.~K. Patel, D.~W. Price, S.~Hatchett, M.~H. Key,
  C.~Andersen, R.~Snavely, and R.~R. Freeman.
\newblock {\em Phys. Rev. Lett.}, 88:215006, 2002.

\bibitem{back_surface}
T.~E. Cowan et~al.
\newblock {\em Phys. Rev. Lett.}, 92:204801, 2004.

\bibitem{structured_7}
H.~Schwoerer, S.~Pfotenhauer, O.~Jackel, K.-U. Amthor, B.~Liesfeld, W.~Ziegler,
  R.~Sauerbrey, K.~W.~D. Ledingham, and T.~Esirkepov.
\newblock {\em Nature}, 439(7075):445--448, 2006.

\bibitem{foam}
I.~Prencipe et~al.
\newblock {\em Plasma Phys. Control. Fusion}, 58(3):034019, 2016.

\bibitem{foam_2}
M.~Passoni et~al.
\newblock {\em Phys. Rev. Accel. Beams}, 19:061301, 2016.

\bibitem{structured_8}
A.~Sgattoni et~al.
\newblock {\em Proc. SPIE}, 8779:87790L--87790L--7, 2013.

\bibitem{absorption_structured}
S.~Zheng-Ming, W.~Su-Ming, Y.~Lu-Le, W.~Wei-Min, C.~Yun-Qian, C.~Min, and
  Z.~Jie.
\newblock {\em Chin. Phys. B}, 24(1):015201, 2015.

\bibitem{absorption_structured_2}
A.~Bigongiari, M.~Raynaud, C.~Riconda, A.~Héron, and A.~Macchi.
\newblock {\em Phys. Plasmas}, 18(10):102701, 2011.

\bibitem{plasmon_4}
M.~Raynaud, J.~Kupersztych, C.~Riconda, J.~C. Adam, and A.~Héron.
\newblock {\em Phys. Plasmas}, 14(9):092702, 2007.

\bibitem{structured_2}
A.~Andreev and K.~Platonov.
\newblock {\em Contrib. Plasma Phys.}, 53(2):173--178, 2013.

\bibitem{structured_11}
S.~Kahaly, S.~K. Yadav, W.~M. Wang, S.~Sengupta, Z.~M. Sheng, A.~Das, P.~K.
  Kaw, and G.~R. Kumar.
\newblock {\em Phys. Rev. Lett.}, 101:145001, 2008.

\bibitem{andreev}
A.~Andreev, K.~Platonov, J.~Braenzel, A.~Lübcke, S.~Das, H.~Messaoudi,
  R.~Grunwald, C.~Gray, E.~McGlynn, and M.~Schnürer.
\newblock {\em Plasma Phys. Control. Fusion}, 58(1):014038, 2016.

\bibitem{nanoSphere}
D.~Margarone et~al.
\newblock {\em Phys. Rev. AB}, 18:071304, 2015.

\bibitem{structured_1}
A.~Andreev, N.~Kumar, K.~Platonov, and A.~Pukhov.
\newblock {\em Phys. Plasmas}, 18(10):103103, 2011.

\bibitem{structured_3}
O.~Klimo, J.~Psikal, J.~Limpouch, J.~Proska, F.~Novotny, T.~Ceccotti,
  V.~Floquet, and S.~Kawata.
\newblock {\em New J. Phys.}, 13(5):053028, 2011.

\bibitem{structured_4}
D.~Margarone et~al.
\newblock {\em Phys. Rev. Lett.}, 109:234801, 2012.

\bibitem{structured_5}
A.~Zigler et~al.
\newblock {\em Phys. Rev. Lett.}, 110:215004, 2013.

\bibitem{structured_6}
A.~Brantov and V.~Bychenkov.
\newblock {\em Contrib. Plasma Phys.}, 53(10):731--735, 2013.

\bibitem{structured_9}
M.~Dalui, W.-M. Wang, T.~M. Trivikram, S.~Sarkar, S.~Tata, J.~Jha, P.~Ayyub,
  Z.~M. Sheng, and M.~Krishnamurthy.
\newblock {\em Sci. Rep.}, 5:11930 EP, 2015.

\bibitem{structured_10}
A.~Andreev, K.~Platonov, J.~Braenzel, A.~L\"{u}bcke, S.~Das, H.~Messaoudi,
  R.~Grunwald, C.~Gray, E.~McGlynn, and M.~Schnürer.
\newblock {\em Plasma Phys. Control. Fusion}, 58(1):014038, 2016.

\bibitem{structured_12}
A.~Zigler et~al.
\newblock {\em Phys. Rev. Lett.}, 106:134801, 2011.

\bibitem{structured_13}
A.~Bigongiari, M.~Raynaud, C.~Riconda, and A.~Héron.
\newblock {\em Phys. Plasmas}, 20(5):052701, 2013.

\bibitem{structured_14}
T.~Ceccotti et~al.
\newblock {\em Phys. Rev. Lett.}, 111:185001, 2013.

\bibitem{plasmon_1}
A.~Sgattoni, L.~Fedeli, G.~Cantono, T.~Ceccotti, and A.~Macchi.
\newblock {\em Plasma Phys. Control. Fusion}, 58(1):014004, 2016.

\bibitem{plasmon_2}
L.~Fedeli et~al.
\newblock {\em Phys. Rev. Lett.}, 116:015001, 2016.

\bibitem{plasmon_3}
J.~Kupersztych, M.~Raynaud, and C.~Riconda.
\newblock {\em Phys. Plasmas}, 11(4):1669--1673, 2004.

\bibitem{plasmon_5}
C.~Riconda, M.~Raynaud, T.~Vialis, and M.~Grech.
\newblock {\em Phys. Plasmas}, 22(7):073103, 2015.

\bibitem{plasmon_6}
A.~Bigongiari, M.~Raynaud, and C.~Riconda.
\newblock {\em Phys. Rev. E}, 84:015402, 2011.

\bibitem{Gibbon}
P.~Gibbon.
\newblock {\em {S}hort {P}ulse {L}aser {I}nteractions with {M}atter: {A}n
  {I}ntroduction}.
\newblock Imperial College Press, London, 2005.

\bibitem{heating}
J.~May, J.~Tonge, F.~Fiuza, R.~A. Fonseca, L.~O. Silva, C.~Ren, and W.~B. Mori.
\newblock {\em Phys. Rev. E}, 84:025401, 2011.

\bibitem{electron_heating}
L.~Cialfi, L.~Fedeli, and M.~Passoni.
\newblock {\em Phys. Rev. E}, 94:053201, 2016.

\bibitem{osiris}
R.~A. Fonseca et~al.
\newblock In {\em Computational Science, ICCS 2002}, volume 2331 of {\em
  Lecture Notes in Computer Science}, pages 342--351. Springer Berlin
  Heidelberg, 2002.

\bibitem{2D3D}
E.~d'Humi\`{e}res, A.~Brantov, V.~Yu. Bychenkov, and V.~T. Tikhonchuk.
\newblock {\em Phys. Plasmas}, 20(2):023103, 2013.

\bibitem{2D3D_2}
A.~Sgattoni, P.~Londrillo, A.~Macchi, and M.~Passoni.
\newblock {\em Phys. Rev. E}, 85:036405, 2012.

\end{thebibliography}

\end{document}